\documentclass[reprint,amsmath,amssymb,aps,prl]{revtex4-2}

\usepackage{graphicx}
\usepackage{dcolumn}
\usepackage{bm}
\usepackage{epstopdf}
\usepackage[mathlines]{lineno}
\usepackage{harpoon}
\usepackage{makecell}
\usepackage{subfigure}
\usepackage{hyperref}

\begin{document}

\title{New Limits on Exotic Spin-Dependent Interactions at Astronomical Distances}

\author{L. Y. Wu}
\affiliation{Key Laboratory of Neutron Physics, Institute of Nuclear Physics and Chemistry, CAEP, Mianyang 621900, Sichuan, China}
\affiliation{Institute of Nuclear Physics and Chemistry, CAEP, Mianyang 621900, Sichuan, China}
\author{K. Y. Zhang}
\affiliation{Key Laboratory of Neutron Physics, Institute of Nuclear Physics and Chemistry, CAEP, Mianyang 621900, Sichuan, China}
\affiliation{Institute of Nuclear Physics and Chemistry, CAEP, Mianyang 621900, Sichuan, China}
\author{M. Peng}
\affiliation{Key Laboratory of Neutron Physics, Institute of Nuclear Physics and Chemistry, CAEP, Mianyang 621900, Sichuan, China}
\affiliation{Institute of Nuclear Physics and Chemistry, CAEP, Mianyang 621900, Sichuan, China}
\author{J. Gong}
\email{gongjian@caep.ac.cn}
\affiliation{Key Laboratory of Neutron Physics, Institute of Nuclear Physics and Chemistry, CAEP, Mianyang 621900, Sichuan, China}
\affiliation{Institute of Nuclear Physics and Chemistry, CAEP, Mianyang 621900, Sichuan, China}
\author{H. Yan}
\email{h.y\_yan@qq.com}
\affiliation{Key Laboratory of Neutron Physics, Institute of Nuclear Physics and Chemistry, CAEP, Mianyang 621900, Sichuan, China}
\affiliation{Institute of Nuclear Physics and Chemistry, CAEP, Mianyang 621900, Sichuan, China}

\date{\today}

\begin{abstract}
  Exotic spin-dependent interactions involving new light particles address key questions in modern physics.
  Interactions between polarized neutrons ($n$) and unpolarized nucleons ($N$) occur in three forms: $g_S^Ng_P^n\boldsymbol{\sigma}\cdot\boldsymbol{r}$, $g_V^Ng_A^n\boldsymbol{\sigma}\cdot\boldsymbol{v}$, and $g_A^Ng_A^n\boldsymbol{\sigma}\cdot\boldsymbol{v}\times\boldsymbol{r}$, where $\boldsymbol{\sigma}$ is the spin and $g$s are the corresponding coupling constants for scalar (S), pseudoscalar (P), vector (V), and axial-vector (A) vertexes.
  If such interactions exist, the Sun and Moon could induce sidereal variations of effective fields in laboratories.
  By analyzing existing data from laboratory measurements on Lorentz and CPT violation, we derived new experimental upper limits on these exotic spin-dependent interactions at astronomical ranges.
  Our limits on $g_S^Ng_P^n$ surpass the previous combined astrophysical-laboratory limits, setting the most stringent experimental constraints to date.
  We also report new constraints on vector-axial-vector and axial-axial-vector interactions at astronomical scales, with vector-axial-vector limits improved by $\sim12$ orders of magnitude.
  We extend our analysis to Hari-Dass interactions and obtain new constraints.
\end{abstract}

\maketitle

{\it Introduction.$-$} Axions, predicted by the PQ (Peccei-Quinn) mechanism~\cite{PEC77,WEI78,WIL78}, can induce spin-dependent interactions~\cite{moody1984new}.
The originally proposed axions were quickly ruled out due to the broken energy scale of the electroweak scale; however, new models with higher broken energy scales were subsequently proposed.
Axions can have arbitrarily small mass and weak couplings to ordinary matter because the scale at which the PQ symmetry is broken can be arbitrarily large~\cite{SVR2006}.
Thus, axions might mediate interactions in ranges from nanometers to astronomical distances.
Though the PQ mechanism was originally proposed to solve the strong CP problem, the axions, which are light, weakly interacting, and pseudoscalar, are also considered as possible candidates for cold dark matter.
New interactions might also be mediated by vector particles such as the para-photon (dark, hidden, heavy or secluded photon)~\cite{HOL1986,DOB2005}, $Z'$ boson~\cite{PDG2020}, graviphoton~\cite{ATW2000}, etc., or even unparticles~\cite{LIA07}. Reference~\cite{dobrescu2006spin} proposed 16 different types of new interactions, 15 of which are spin-dependent.
Non-Yukawa exotic interactions due to the dark or hidden sector were also proposed recently~\cite{COS2020,BRA2020}.
As early as 1980, Fayet~\cite{FAY1980a,FAY1980b} pointed out that the new U(1) vector bosons,characterized by small masses and weak couplings to ordinary matter, could be generated through the spontaneous breaking of supersymmetric theories.
Searching for the new interactions mediated by the new particles is related to the strong CP problem, dark matter, dark energy, and finding evidence of supersymmetry, which is among the most important unsolved problems in modern physics~\cite{RMP18}.

Particles with similar properties as axions, predicted by various generalized theories, are usually called ALPs (Axion Like Particles)~\cite{PDG2020}.
If exist, ALPs ($\phi$) can generate a new interaction through the coupling to a fermion $\psi$, $\mathcal{L}_{\phi}=\bar{\psi}(g_{S}+ig_{P}\gamma_{5})\psi\phi$, where $g_S$ ($g_P$) is the scalar (pseudoscalar) coupling constant~\cite{moody1984new}.
The SP (scalar-pseudoscalar) interaction between a polarized neutron and an unpolarized nucleon can be expressed as
\begin{equation}\label{eqnSP}
V_{\rm{SP}}=\frac{\hbar^2g^N_Sg^n_P}{8\pi m_n}(\frac{1}{\lambda r}+\frac{1}{r^2}){\rm{exp}}(-r/\lambda)\boldsymbol{\sigma}\cdot\hat{r},
\end{equation}
where $\lambda=\hbar/m_{\phi}c$ is the interaction range, $m_{\phi}$ is the mediator mass, $\boldsymbol{\sigma}$ is the spin operator of the polarized neutron, $m_n$ is the neutron mass, $N$ ($n$) represents the nucleon (neutron), and $r$ is the distance between the two interacting particles.
This interaction could also be generated from $\mathcal{L}{\phi}=g_{S}\bar{\psi}\psi\phi+g_{P}/(2m)\bar{\psi}\gamma_{\mu}\gamma_{5}\psi\partial^{\mu}\phi$~\cite{moody1984new,raffelt1996stars,OHA2020}, which includes a derivative term.
Such a term appears in the axion models~\cite{BOOK1996}.
Notably, the SP interaction generated from the derivative coupling between axions and fermions is discussed in Ref.~\cite{RMP2021}.
Thus, studying this interaction can be used to investigate not only ALPs but also axions.
Recently, the SP interaction has begun to attract more attention~\cite{ARV2014,TER2015,CRE2017,GER17,LEE2018,AGG20,CRE2022}.
For example, Wei \emph{et al}.~\cite{WEI2023} proposed a laboratory experiment scheme that could surpass the astrophysical limit for the SP interaction.
Laboratory limits become closer and closer to the limits derived by combining $g_S^N$ from the torsion balance experiment and $g_P^n$ from SN1987A; however, all the existing limits are less stringent than the combined astrophysical-laboratory ones at astronomical distances.

VA (vector-axial-vector) and AA (axial-axial-vector) interactions can be derived from a general Lagrangian $\mathcal{L}_{X}=X_\mu\bar{\psi}(g_V\gamma^\mu+g_A\gamma_5\gamma^\mu)\psi$ in the nonrelativistic limit,
where $X$ is a new vector particle and $g_V$ ($g_A$) is the vector (axial-vector) coupling constant.
The interactions between a polarized neutron and an unpolarized nucleon can be expressed as
\begin{eqnarray}
V_{\rm{VA}}&=&\frac{\hbar g^N_Vg^n_A}{2\pi}\frac{{\rm{exp}}(-r/\lambda)}{r}\boldsymbol{\sigma}\cdot\boldsymbol{v}, \label{md} \\
V_{\rm{AA}}&=&\frac{\hbar^2g^N_Ag^n_A}{16\pi m_nc}(\frac{1}{\lambda r}+\frac{1}{r^2}){\rm{exp}}(-r/\lambda)\boldsymbol{\sigma}\cdot(\boldsymbol{v}\times\hat{r}), \label{md2}
\end{eqnarray}
where $\boldsymbol{v}$ is their relative velocity.
Many studies have been carried out to look for the new interactions, detecting either the macroscopic forces or the torques exerted on the polarized probe spins.
For example, Leslie {\em et al}.~\cite{LES2014} proposed experimental schemes to detect the new spin-dependent interaction between a polarized source and a mechanical oscillator.
For another example, Ding {\em et al}.~\cite{Ding2020} used a micro-fabricated magnetic structure as a polarized source, and then tried to detect the AA interaction at the range of $\sim\mu$m sensed by a gold-sphere-cantilever.
Many groups have been searching for the new interaction through its rotating effects as a effective magnetic field on the polarized spin.
The VA and AA interactions between different combinations of fermions have been investigated already, such as electron-nucleon~\cite{Kim2018,Kim2019,Ji2018,Ding2020}, neutron-nucleon~\cite{Piegsa2012,Yan2013,Yan2015}, electron-electron~\cite{Ficek2017}, and electron-antiproton~\cite{Ficek2018}.
Studies on these exotic interactions involving muons were performed very recently~\cite{YAN19}.

Since the signal induced by new interactions is tiny, using a large mass source and modulating the signal to a high frequency are crucial for the detection.
A mass source with vast constituents could make the signal measurable.
By modulating the signal, on the one hand, the SNR (Signal-to-Noise Ratio) can be increased with the decrease of the noise bandwidth.
On the other hand, the $1/f$ noise can be significantly reduced at high frequencies.
In this Letter, treating the Sun and the Moon as mass sources and using the Earth's rotation as a modulation, we obtain new limits on exotic spin-dependent SP (\ref{eqnSP}), VA (\ref{md}), and AA (\ref{md2}) interactions at astronomical distances.

{\it The basic idea.$-$} All three types of spin-dependent interactions are in the form of $\boldsymbol{s}\cdot\boldsymbol{B}'$, where $\boldsymbol{B}'$ can be viewed as a kind of effective magnetic field.
Searching for these interactions becomes a problem probing the effective field acting on polarized spins.
\begin{figure*}[htbp]
	\centering
	\subfigure[]{\includegraphics[width=0.45\textwidth]{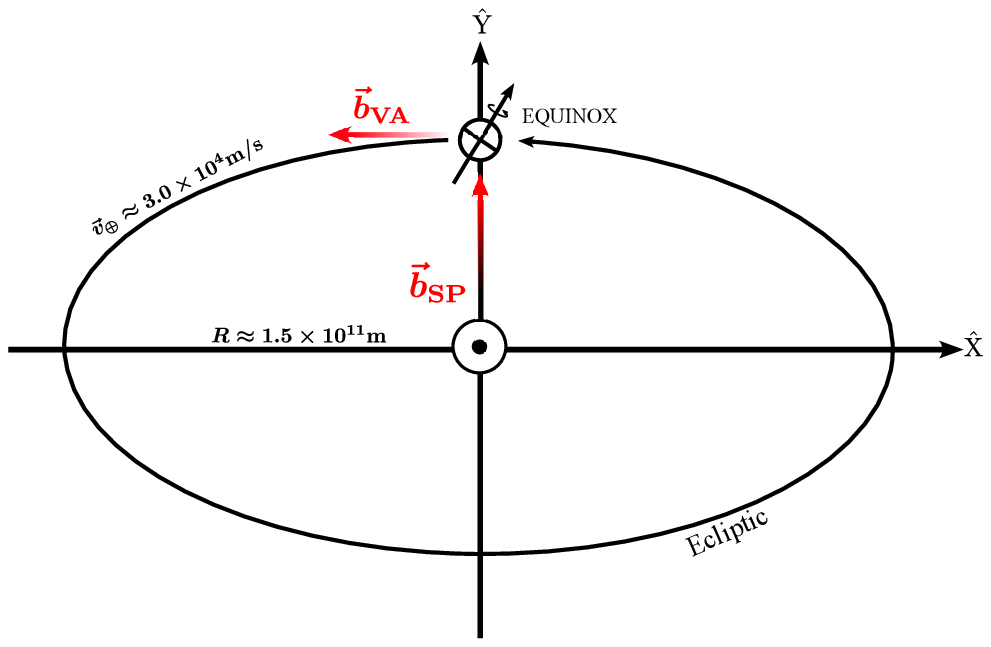}}
	\subfigure[]{\includegraphics[width=0.45\textwidth]{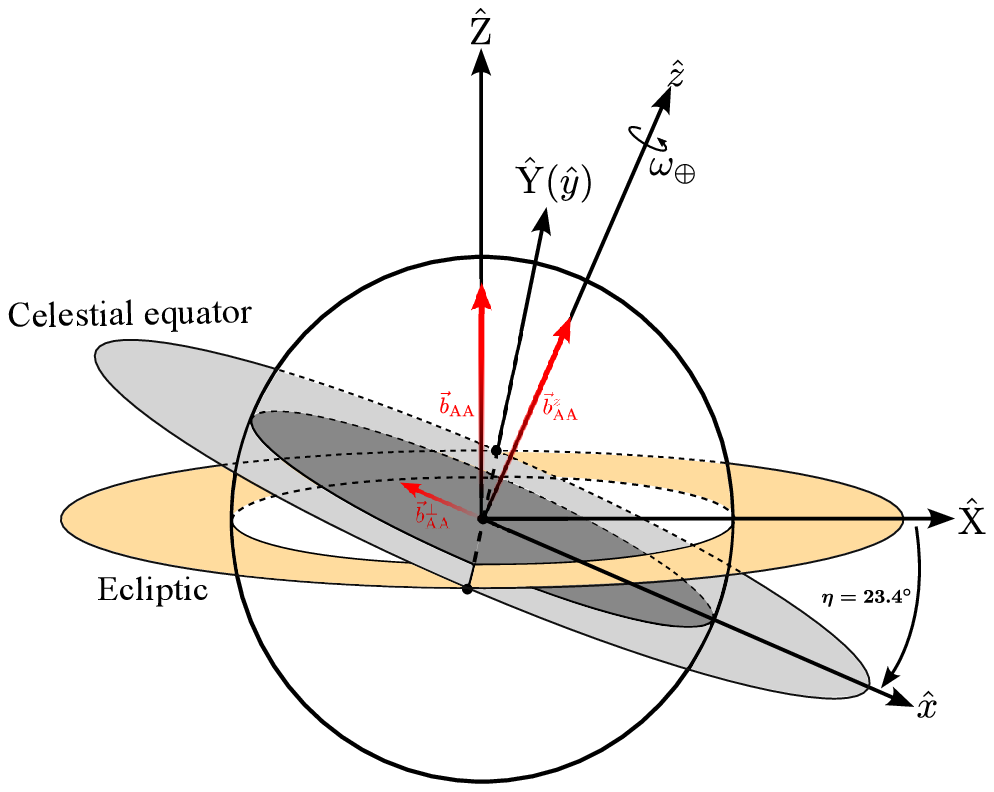}}
	\caption{(a) The Sun-centered frame. The relative size of the Sun and the Earth is not to scale. (b) The Earth-based frame. We take $\hat{z}$ along with the Earth's rotation axis. The angle between the ecliptic plane and the Earth's equatorial plane is $\eta=23.4^\circ$. The red arrows represent the directions of effective fields of three types of spin-dependent interactions generated by the Sun's nucleons.}
	\label{CT}
\end{figure*}
We first illustrate the basic idea using the Sun as the source mass.
In the Sun-centered frame shown in Fig.~\ref{CT} (a), the aforementioned new interactions can generate effective magnetic fields at the Earth center as \cite{SUPP}
\begin{widetext}
\begin{equation}
\begin{split}
\boldsymbol{B}'_{\rm{SP}}&=\frac{\hbar g^N_Sg^n_PN_\odot}{4\pi m_n\gamma_n}(\frac{1}{\lambda R}+\frac{1}{R^2}){\rm{exp}}(-R/\lambda)[\cos{(\Omega_\oplus t)}\hat{X}+\sin{(\Omega_\oplus t)}\hat{Y}], \\
\boldsymbol{B}'_{\rm{VA}}&=\frac{g^N_Vg^n_AN_\odot}{\pi\gamma_n}\frac{{\rm{exp}}(-R/\lambda)}{R}[-\Omega_\oplus R\sin{(\Omega_\oplus t)}\hat{X}+\Omega_\oplus R\cos{(\Omega_\oplus t)}\hat{Y}], \\
\boldsymbol{B}'_{\rm{AA}}&=-\frac{\hbar g^N_Ag^n_AN_\odot}{8\pi m_nc\gamma_n}(\frac{1}{\lambda R}+\frac{1}{R^2}){\rm{exp}}(-R/\lambda)\Omega_\oplus R\hat{Z},	
\end{split}
\end{equation}
\end{widetext}
where $R$ is the distance from the Earth to the Sun, $\gamma_n$ is the gyromagnetic ratio of the neutron, $\Omega_\oplus$ is the Earth's orbital angular frequency, and $N_\odot$ is the total nucleon number of the Sun.
For a laboratory frame on the Earth as shown in Fig.~\ref{CT} (b), the effects of the Earth's rotation can be taken into account via the Euler rotations. We first rotate the frame by the angle $\omega_\oplus t$ about the $\hat{Z}$ axis where $\omega_\oplus$ is the Earth's rotation frequency, and then rotate about the $\hat{Y}$ axis by an angle $\eta$ which is the Earth's obliquity.
In the laboratory frame, we will observe effective time-varying fields as
\begin{widetext}
\begin{align}
	&\boldsymbol{b}_{\rm{SP}}=\frac{\hbar g^N_Sg^n_PN_\odot}{4\pi m_n\gamma_n}(\frac{1}{\lambda R}+\frac{1}{R^2}){\rm{exp}}(-R/\lambda)\Bigg [\begin{array}{c}
		\cos{\eta}\cos{(\Omega_\oplus t)}\cos{(\omega_\oplus t)}+\sin{(\Omega_\oplus t)}\sin{(\omega_\oplus t)}\\
		-\cos{\eta}\cos{(\Omega_\oplus t)}\sin{(\omega_\oplus t)}+\sin{(\Omega_\oplus t)}\cos{(\omega_\oplus t)}\\
		\sin{\eta}\cos{(\Omega_\oplus t)}
		\end{array}\Bigg ], \label{bsp}\\
	&	\boldsymbol{b}_{\rm{VA}}=\frac{g^N_Vg^n_AN_\odot}{\pi\gamma_n}\frac{{\rm{exp}}(-R/\lambda)}{R}v_\oplus\Bigg [\begin{array}{c}
            - \cos{\eta}\cos{(\omega_\oplus t)}\sin{(\Omega_\oplus t)}+\sin{(\omega_\oplus t)}\cos{(\Omega_\oplus t)}\\
            \cos{\eta}\sin{(\omega_\oplus t)}\sin{(\Omega_\oplus t)}+\cos{(\omega_\oplus t)}\cos{(\Omega_\oplus t)}\\
            -\sin{\eta}\sin{(\Omega_\oplus t)}
        \end{array}\Bigg ],\label{bva} \\
	&\boldsymbol{b}_{\rm{AA}}=\frac{\hbar g^N_Ag^n_AN_\odot}{8\pi m_nc\gamma_n}(\frac{1}{\lambda R}+\frac{1}{R^2}){\rm{exp}}(-R/\lambda)v_\oplus\Bigg [\begin{array}{c}
		\sin{\eta}\cos{(\omega_\oplus t)}\\
		-\sin{\eta}\sin{(\omega_\oplus t)}\\
	  -\cos{\eta}
   \end{array}\Bigg ] \label{baa},
\end{align}
\end{widetext}
where $v_\oplus=\Omega_\oplus R$  is the orbital speed of the Earth.
As the most straightforward case, $ \boldsymbol{b}_{\rm{AA}}$ clearly shows effective magnetic fields rotating in the laboratory frame at the Earth's rotation frequency.
Although $\boldsymbol{b}_{\rm{SP}}$ and $\boldsymbol{b}_{\rm{VA}}$ appear more complicated due to their mixture with the Earth's orbital rotation, the situation can be greatly simplified by considering the fact that $\omega_\oplus\gg \Omega_\oplus$.

In summary, if the exotic spin-dependent interactions exist, the perpendicular components of the effective fields induced by the Sun are modulated by the Earth's rotations; thus, we could observe its signal in the laboratory.
Although the frequency of the Earth's rotation is not high, its frequency modulation effect on the nuclear precession in the comagnetometer makes
precision measurements on these new interactions possible.

{\it Constraining the exotic spin-dependent interactions at astronomical distances.$-$}Dual-species comagnetometers are convenient for detecting the tiny signals caused by new spin-dependent interactions, since the two components occupy the same space and common-mode background field noise can be mostly canceled.
In principle, we can separate the sidereal modulated signal from the noisy background during precise measurements.
The ultrahigh sensitivity of the comagnetometer to the magnetic field changes has an extensive implementation in new physics detection, including electric dipole moments, CPT and Lorentz violation, spin-gravity interaction, and so on~\cite{sachdeva2019new,venema1992search,chupp2019electric,alonso2019exploring}.
This comagnetometer method has been used to search for the constant cosmic background field due to Lorentz violation by detecting the sidereal variants of the field observed in the laboratory frame on the Earth in Refs.~\cite{allmendinger2014new,brown2010new}, where a $^{129}$Xe+$^{3}$He comagnetometer and a K+$^3$He one were respectively adopted.
Stringent constraints on the components of the constant field perpendicular to the Earth's rotation axis at a similar level were obtained.

We find that the limits on exotic spin-dependent interactions induced by the Sun can be obtained by using the Lorentz violation searching results.
For example, for the experiment described in Ref.~\cite{allmendinger2014new}, the $\Omega_\oplus t$ in Eqs.~(\ref{bsp}), (\ref{bva}) and (\ref{baa}) approximately equals to $\pi/2$, given that the experiment was performed for $\sim$10 days when the Earth was around the vernal equinox.
The sidereal oscillating effective field $\boldsymbol{b}_{\perp}$ perpendicular to the Earth's rotation axis can be detected as~\cite{SUPP}
\begin{equation}
 \begin{aligned}
 \boldsymbol{b}_{\rm{SP}\perp}=&\frac{\hbar g^N_Sg^n_PN_\odot}{4\pi m_n\gamma_n}(\frac{1}{\lambda R}+\frac{1}{R^2})\exp{(-R/\lambda)}\\
 &[\sin{(\omega_\oplus t)}\hat{x}+\cos{(\omega_\oplus t)}\hat{y}],\\
 \boldsymbol{b}_{\rm{VA\perp}}=&-\frac{g^N_Vg^n_AN_\odot}{\pi\gamma_n}\frac{{\rm{exp}}(-R/\lambda)}{R}v_\oplus\cos{\eta}\\
 &[-\cos{(\omega_\oplus t)}\hat{x}+\sin{(\omega_\oplus t)}\hat{y}],\\
 \boldsymbol{b}_{\rm{AA\perp}}=&\frac{\hbar g^N_Ag^n_AN_\odot}{8\pi m_nc\gamma_n}(\frac{1}{\lambda R}+\frac{1}{R^2}){\rm{exp}}(-R/\lambda)v_\oplus\sin{\eta}\\
 &[\cos{(\omega_\oplus t)}\hat{x}-\sin{(\omega_\oplus t)}\hat{y}].
\end{aligned}
\end{equation}
Using the result of Ref.~\cite{allmendinger2014new}, at the $95\%$ confidence level ($95\%$ C.L.), we could derive
\begin{equation}
|\boldsymbol{b}_\perp|<0.023~\rm{fT}.
\end{equation}
Plugging in all the known parameters such as $\eta=23.4^\circ$, $N_\odot\approx 1.2\times 10^{57}$, $R=1.5\times 10^{11}$~m, and $v_\oplus\approx 3.0\times 10^{4}$~m/s, we can obtain the constraints on the SP (\ref{eqnSP}), VA (\ref{md}), and AA (\ref{md2}) interactions between polarized neutrons and unpolarized nucleons.

\begin{figure}[htbp]
\centering
\includegraphics[width=0.45\textwidth]{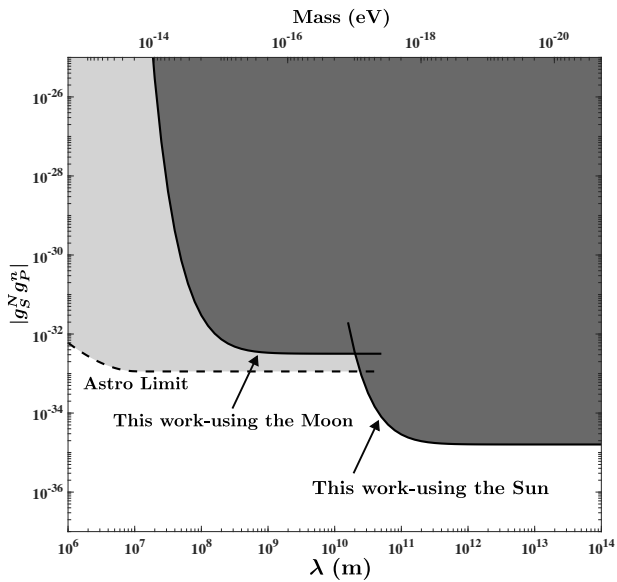}
\caption{Constraint to the coupling constant product $|g^N_Sg^n_P|$ as a function of the interaction range $\lambda$ (the mediator mass). The solid lines are the result of this work; the left line uses the Moon as the source, and the right line uses the Sun. The dashed line is the result of Refs.~\cite{raffelt2012limits,OHA2020}, which was derived by combining $g_S^N$ of weak equivalence with $g_P^n$ from SN1987A. The dark grey area is excluded by the result of this work and the light gray one is excluded by the result of Refs.~\cite{raffelt2012limits,OHA2020}.}
\label{gsgp}
\end{figure}

The derived constraint on $|g_S^Ng_P^n|$ is shown in Fig.~\ref{gsgp}.
For $\lambda\gtrsim 2\times10^{10}$~m, it gives the most stringent limit on $|g_S^Ng_P^n|$.
For $\lambda>10^{12}$~m,  our bounds $|g_S^Ng_P^n|<1.6\times10^{-35}$ ($95\%$ C. L.).
Previously, the most stringent constraints on $|g_S^Ng_P^n|$ were astrophysical-laboratory limits that combined astrophysical constraints on $g_P^n$ from SN1987A with the laboratory ones on $g_S^N$ from the weak equivalence principle experiment.
This work improves the existing upper bound by as much as $\sim$70 times.
In particular, the new limits on the scalar-pseudoscalar coupling combination exceed the combined astrophysical-laboratory limits for the first time.

\begin{figure}[htbp]
\centering
\includegraphics[width=0.45\textwidth]{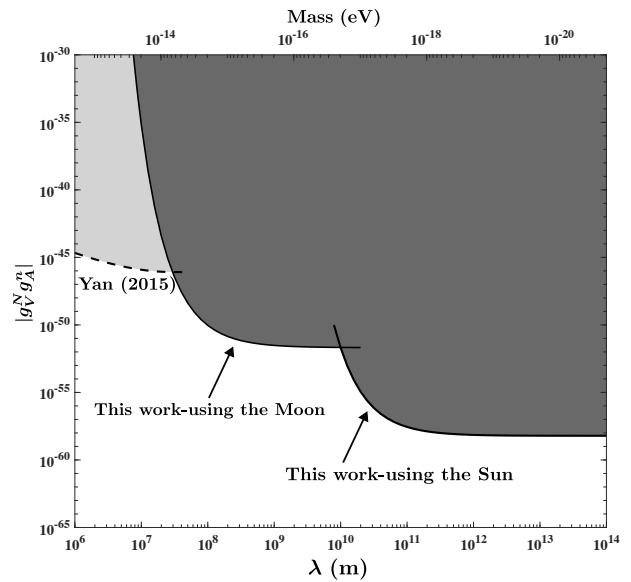}
\caption{Constraint to the coupling constant product $|g^N_Vg^n_A|$ as a function of the interaction range $\lambda$ (the mediator mass).  The solid lines are the result of this work; the left line uses the Moon as the source, and the right line uses the Sun. The dashed line is the result of Ref.~\cite{Yan2015}. The dark grey area is excluded by the result of this work and the light gray one is excluded by the result of Ref.~\cite{Yan2015}.}
\label{gvga}
\end{figure}

Our constraint on $|g^N_Vg^n_A|$ is shown in Fig.~\ref{gvga}.
For $\lambda\gtrsim 3\times10^{7}$~m, it gives the most stringent limit on $|g^N_Vg^n_A|$.
For $\lambda>10^{12}$~m,  our bounds $|g^N_Vg^n_A|<7.1\times10^{-59}$ ($95\%$ C. L.).
Previously, the most stringent constraint on  $|g^N_Vg^n_A|$ near the interaction range under consideration was given in Ref.~\cite{Yan2015}.
If the previous result can be extended to the range of $\sim 10^{12}$~m, the present work improves the existing upper bound by as much as 12 orders of magnitude.

\begin{figure}[htbp]
\centering
\includegraphics[width=0.45\textwidth]{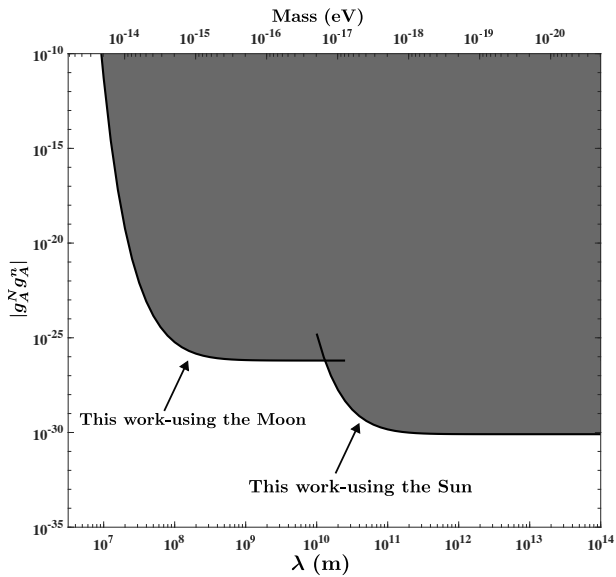}
\caption{Constraint to the coupling constant product $|g^N_Ag^n_A|$ as a function of the interaction range $\lambda$ (the mediator mass).  The solid lines are the result of this work; the left line uses the Moon as the source, and the right line uses the Sun. The dark grey area is excluded by the result of this work.}
\label{gaga}
\end{figure}

The obtained constraint on $|g_A^Ng_A^n|$ is shown in Fig.~\ref{gaga}.
For $\lambda\gtrsim10^{7}$~m, it gives the first limit on $|g_A^Ng_A^n|$.
For $\lambda>10^{12}$~m,  our bounds $|g_A^Ng_A^n|<8.1\times10^{-31}\rm{(95\%~{\mathrm C. L.})}$.
It is the only known constraint on $|g_A^Ng_A^n|$ to us at the astronomical ranges.

We can apply the same analyzing method by using the Moon as a source. In this case, we shall consider errors due to several systematic effects. We show details of the analysis in the Supplementary Material \cite{SUPP}. We also plotted the results using the Moon in Figs.~\ref{gsgp},~\ref{gvga}, and ~\ref{gaga}.

Furthermore, the spin-gravity interaction proposed by Leitner and Okubo~\cite{leitner1964parity} and later generalized by Hari Dass~\cite{dass1976test} can also be strictly constrained.
Assuming CPT invariance, two types of discrete symmetry violation spin-gravity potentials are constructed as
\begin{equation}
\label{SG}
V(r)=\frac{G_NM\hbar}{2}(\alpha_1\frac{\boldsymbol{\sigma}\cdot\hat{r}}{cr^2}+\alpha_2\frac{\boldsymbol{\sigma}\cdot\boldsymbol{v}}{c^2r^2}),
\end{equation}
where $G_N$ is the Newton constant of gravitation, $M$ is the mass of the gravity source, and $\alpha_1$ ($\alpha_2$) is a dimensionless constant.
These potentials are the starting point of many low-energy experiments~\cite{bargueno2008constraining,kimball2017constraints}.
They also provide a direct way to test symmetry violation and the equivalence principle in General Relativity~\cite{dass1976test,duan2016test}.
Using the Sun as the mass source, we derive the limits on $\alpha$ as
\begin{equation}
\begin{split}
&|\alpha_1|<2.2\times 10^2~\rm{(95\%~{\mathrm C. L.})},\\
&|\alpha_2|<2.4\times 10^6~\rm{(95\%~{\mathrm C. L.})}.
\end{split}
\end{equation}
When comparing with the results of Ref.~\cite{altarev2009test}, our limit on $\alpha_1$ improves the existing one by $\sim$11 times, and on $\alpha_2$ we get an improvement of $\sim$4 orders of magnitude.

{\it Conclusion and discussion.$-$}By using the Sun and the Moon as sources, the Earth's rotation as modulation, and the existing laboratory limits on the Lorentz and CPT violation  at distances of astronomical scales, we have constrained three types of possible new interactions between polarized neutrons and unpolarized nucleons.
We derived new limits on the SP interaction with ranges from $\sim2\times10^{10}$~m to $\sim10^{14}$~m.
At the distance of $\sim10^{12}$~m, the limit is improved by $\sim$70 times compared to the previous combined astrophysical-laboratory limit.
This result is the first time the limits from a single laboratory experiment exceed the combined astrophysical-laboratory ones for the SP interaction.
We obtained new limits on the VA interaction with ranges from $\sim3\times10^{7}$~m to $\sim10^{14}$~m.
At the distance of $\sim10^{12}$~m, the limit is improved by $\sim$12 orders of magnitude in comparison with the previous result of $^3$He spin relaxation experiment.
We derived the first limits on the AA interaction with ranges from $\sim10^{7}$~m to $\sim10^{14}$~m.
We also constrained the Hari-Dass spin-dependent interactions and obtained new limits on them.

How can we extend the current work to include other particles, such as electrons and muons?
One possibility is to employ the beam method proposed in Refs.~\cite{YAN14,YAN19}, which uses superconducting magnetic shielding to create a region with zero background field.
By directing spin-polarized particle beams through this region and detecting sidereal variations in polarization along the direction perpendicular to the Earth's rotation axis, we can investigate spin-dependent new interactions induced by the Sun for the probing electrons, muons, etc.

{\it Note added.$-$}After this work was submitted, Ref. \cite{Zhang2023arXiv} reports improved limits on $g_S^Ng_P^n$ in the interaction range of $10^6\sim10^{10}~\mathrm{m}$, and Ref. \cite{clayburn2023} reports similar limits on $g_A^Ng_A^n$ in the interaction range of $\sim10^{12}~\mathrm{m}$.

\begin{acknowledgments}
We acknowledge supports from the National Natural Science Foundation of China under grants U2230207 and U2030209.
This work was also supported by the National Key Program for Research and Development of China under grants 2020YFA0406001 and 2020YFA0406002.
We thank Dr. C. Fu and Y. M. Ma for their helpful discussions.
\end{acknowledgments}

\end{document}